\documentclass[sigconf,screen,9pt]{acmart}

\def\conference
\usepackage{etoolbox}
\newcommand{\ifconference}[1]{{{\ifx\fullversion\undefined{#1}\fi}\xspace}}
\newcommand{\iffullversion}[1]{{{\ifx\conference\undefined{#1}\fi}\xspace}}

\usepackage{graphicx}  % pictures and figures
\usepackage{lipsum}  % random paragraphs
\newcommand{\hide}[1]{} % hide
\usepackage{xspace}
\usepackage{textcomp}
\usepackage{comment} % \begin{comment} ... \end{comment}
\usepackage{verbatim}
\usepackage{url}

\newcommand{\defn}[1]{\emph{\textbf{#1}}} % definition style
\newcommand{\emp}[1]{\emph{\textbf{#1}}} % highlight
\let \originalleft \left
\let\originalright\right
\renewcommand{\left}{\mathopen{}\mathclose\bgroup\originalleft}
\renewcommand{\right}{\aftergroup\egroup\originalright}

\usepackage{scalerel} % stretch a formula

\newtheoremstyle{exampstyle}
{.5em} % Space above
{1em} % Space below
{\it} % Body font
{.5em} % Indent amount
{\it \bfseries} % Theorem head font
{.} % Punctuation after theorem head
{.5em} % Space after theorem head
{} % Theorem head spec (can be left empty, meaning `normal')
\theoremstyle{exampstyle} 
\theoremstyle{exampstyle} 
\theoremstyle{exampstyle} 
\theoremstyle{exampstyle} 

\makeatletter

\newcommand{\whp}[1]{\emph{whp}}

\usepackage{pifont}
\newcommand{\modelop}[1]{\texttt{#1}}
\newcommand{\forkins}{\modelop{fork}}

\newcommand{\thread}{thread}

\usepackage[shortlabels]{enumitem}

\setlist{topsep=0.3em,itemsep=0.2em,parsep=0.1em,leftmargin=*}
\usepackage{float}
\usepackage[labelfont=bf,font={small},aboveskip=0em, belowskip=0em]{caption}

\usepackage[labelfont=bf,list=true,skip=0em]{subcaption}
\usepackage[rightcaption]{sidecap}

\usepackage{wrapfig}

\usepackage{array}% for extended column definitions
\newcolumntype{L}[1]{>{\raggedright\let\newline\\\arraybackslash\hspace{0pt}}m{#1}}
\newcolumntype{C}[1]{>{\centering\let\newline\\\arraybackslash\hspace{0pt}}m{#1}}
\newcolumntype{R}[1]{@{}>{\ }r<{\ }@{}}
\newcolumntype{B}{>{\bf}c}
\usepackage{rotating}

\usepackage{booktabs} % provides toprule, bottomrule, midrule, cmidrule, etc.
\usepackage{multicol,multirow}
\usepackage{longtable} % provides long table
\usepackage{supertabular} % similar to long table, allowing tables to take more than one page
\usepackage{colortbl}
\usepackage{bigstrut}
\usepackage{titlesec}
\titlespacing{\section}{0pt}{0.3em}{0.2em} % left margin, space before, space after
\titlespacing{\subsection}{0pt}{0.3em}{0.2em} % left margin, space before, space after
\titlespacing{\subsubsection}{0pt}{0.1em}{1em} % left margin, space before, space after (horizontal)
\newcommand{\mysubsubsection}[1]{\underline{#1}.}
\titleformat{\subsubsection}[runin]
{\normalfont\normalsize\bfseries}{\thesubsubsection}{1em}{\mysubsubsection}

\newcommand{\myparagraph}[1]{\noindent\emp{#1}}

\usepackage[ruled,lined,linesnumbered,noend]{algorithm2e}
\usepackage[noend]{algpseudocode}

\makeatletter
\patchcmd{\@algocf@start}% <cmd>
  {-1.5em}% <search>
  {0pt}% <replace>
  {}{}% <success><failure>
\newcommand{\nosemic}{\renewcommand{\@endalgocfline}{\relax}}% Drop semi-colon ;
\newcommand{\dosemic}{\renewcommand{\@endalgocfline}{\algocf@endline}}% Reinstate semi-colon ;
\SetSideCommentLeft
\SetKwInput{notations}{Notations}
\SetKwInput{notes}{Notes}
\SetKwInput{maintains}{Maintains}

\SetKwProg{myfunc}{Function}{}{}
\SetKwFor{parForEach}{ParallelForEach}{do}{endfor}
\SetKwFor{Justrepeat}{Repeat}{}{}

\SetKw{MIN}{min}
\SetKw{MAX}{max}
\SetKw{OR}{or}
\SetKw{AND}{and}

\SetCommentSty{mycommfont}

\usepackage{mdframed}
\definecolor{framelinecolor}{RGB}{68,114,196}
\mdfdefinestyle{mystyle}{linecolor=framelinecolor,innertopmargin=1pt,innerbottommargin=2pt,backgroundcolor=gray!20,skipabove=2pt,skipbelow=0pt}%leftmargin=0,topmargin=
\mdfdefinestyle{densestyle}{linecolor=framelinecolor,innertopmargin=0,innerbottommargin=0,leftmargin=0,rightmargin=0,backgroundcolor=gray!20}
\mdfdefinestyle{compactcode}{linecolor=framelinecolor,innertopmargin=1pt,innerbottommargin=1pt,backgroundcolor=gray!20,skipabove=0pt,skipbelow=0pt,leftmargin=0,rightmargin=0}

\usepackage{framed}

\usepackage{listings}

\newdimen\zzsize
\newdimen\kwsize
\newcommand{\basicstyle}{\fontsize{\zzsize}{1\zzsize}\ttfamily}
\newcommand{\keywordstyle}{\fontsize{\kwsize}{1\kwsize}\ttfamily\bf}

\newdimen\zzlstwidth
\usepackage{cleveref}
\crefname{section}{Sec.}{Sec.}
\crefname{theorem}{Thm.}{Thm.}
\crefname{lemma}{Lem.}{Lem.}
\crefname{corollary}{Col.}{Col.}
\crefname{table}{Tab.}{Tab.}
\crefname{algorithm}{Alg.}{Alg.}
\crefname{figure}{Fig.}{Fig.}
\crefname{fact}{Fact}{Fact}
\Crefname{table}{Tab.}{Tab.}
\crefname{problem}{Problem}{Problem}

\usepackage{tikz} % draw geometric objects

\setcopyright{none}
\renewcommand\footnotetextcopyrightpermission[1]{} % This line removes the footnote about the conference and year.
\newcommand{\pasgal}{PASGAL}
\newcommand{\gbbs}{GBBS}
\newcommand{\gapbs}{GAPBS}

\newcommand{\ff}{\mathcal{F}}

\newcommand{\knn}{$k$-NN}

\newcommand{\lowdiam}{low-diameter}
\newcommand{\largediam}{large-diameter} 

\AtBeginDocument{%
  }

\copyrightyear{2024}
\acmYear{2024}
\setcopyright{rightsretained}
\acmConference[SPAA '24]{Proceedings of the 36th ACM Symposium on Parallelism in Algorithms and Architectures}{June 17--21, 2024}{Nantes, France}
\acmBooktitle{Proceedings of the 36th ACM Symposium on Parallelism in Algorithms and Architectures (SPAA '24), June 17--21, 2024, Nantes, France}\acmDOI{10.1145/3626183.3660258}
\acmISBN{979-8-4007-0416-1/24/06}

\begin{document}

\title{PASGAL: Parallel And Scalable Graph Algorithm Library}
%\shorttitle{}

\settopmatter{authorsperrow=4}
\author{Xiaojun Dong}
\affiliation{\institution{UC Riverside}\city{}\country{}}
\email{xdong038@ucr.edu}
\author{Yan Gu}
\affiliation{\institution{UC Riverside}\city{}\country{}}
\email{ygu@cs.ucr.edu}
\author{Yihan Sun}
\affiliation{\institution{UC Riverside}\city{}\country{}}
\email{yihans@cs.ucr.edu}
\author{Letong Wang}
\affiliation{\institution{UC Riverside}\city{}\country{}}
\email{lwang323@ucr.edu}



\begin{abstract}
  In this paper, we introduce \pasgal{} (Parallel And Scalable Graph Algorithm Library), 
  a parallel graph library that scales to a variety of graph types, many processors, 
  and large graph sizes. One special focus of PASGAL is the efficiency on \emph{large-diameter graphs},
  which is a common challenge for many existing parallel graph processing systems: 
  many existing graph processing systems can be even slower than the
  standard sequential algorithm on large-diameter graphs due to the lack of parallelism. 
  \hide{This issue occurs on many graph problems, even some fundamental ones, such as breadth-first-search (BFS),
  strongly connected components (SCC), biconnected components (BCC), single-source shortest-paths (SSSP), etc.}
  Such performance degeneration is caused by the high overhead in scheduling and synchronizing threads when traversing
  the graph in the breadth-first order. 
    
  \hide{Such performance degeneration is caused by traversing
  the graph from one (sometimes multiple) vertex in rounds, with thread synchronization needed between every two rounds. 
  When the diameter of the graph is large,
  many rounds are required, making this scheduling overhead more expensive than the computation.}
  
  The core technique in \pasgal{} to achieve high parallelism is a technique called \emph{vertical granularity control (VGC)} to hide synchronization overhead, as
  well as careful redesign of parallel graph algorithms and data structures. 
  %Inspired by standard granularity control in parallel programming, 
  %VGC hides the scheduling overhead by locally searching a reasonable large neighborhood when visiting a vertex in traversal.
  %In this way, algorithms in \pasgal{} require fewer rounds to finish, and enable much higher parallelism and better scalability. 
  In our experiments, we compare \pasgal{} with state-of-the-art parallel implementations on BFS, SCC, BCC, and SSSP.
  \pasgal{} achieves competitive performance on small-diameter graphs compared to the parallel baselines, 
  and is significantly faster on large-diameter graphs. 
  
  \hide{To improve the scalability, \pasgal{} employs several techniques,
  including algorithm redesign, 
  hiding scheduling overhead using \emp{vertical granularity control (VGC)},
  and using new parallel data structures. 
  %In this way, algorithms in \pasgal{} achieve high parallelism and good scalability. 
  In our experiments, %we compare \pasgal{} with state-of-the-art parallel implementations.
  \pasgal{} achieves competitive performance on small-diameter graphs compared to the baselines, and is much faster on large-diameter graphs. }
  \hide{
  The first is to redesign certain algorithms to reduce the number of .
  Second, we employ a new technique in recent papers, called \emph{vertical granularity control (VGC)}, to hide the scheduling overhead and 
  reduce number of global synchronizations.
  Finally, several new data structures are introduced, to maintain sets of vertices with dynamic updates more efficiently. 
  }
  
\end{abstract}




\fancyhead{} % This line removes the page headers about the conference and authors.

\maketitle

\section{Introduction} 

Graphs are effective representations of real-world objects and their relationships. 
%such as social network users and their friendship,
%webpages and their links, geological objects and roads between them, machine learning objects and their similarities, etc. 
Processing and analyzing graphs efficiently have become increasingly important.
%One of the most important recent hardware advance that has impacted graph processing is the popularization of \emph{multicore systems.}
Given the large size of today's real-world graphs, it is imperative to consider parallelism in graph processing.
The increasing number of cores and memory size
allows a single machine to easily process graphs with billions of vertices in a few seconds,
even for reasonably complicated tasks.
As a result, a huge number of in-memory graph processing algorithms and systems have been developed (e.g.~\cite{gbbs2021,beamer2015gap}).
%Such parallel systems are applicable on from desktop PC (usually 4-16 processors) to high-end servers (hundreds of processors or more),
%and the number of core keeps increasing. 

\begin{figure}
  \centering
  \includegraphics[width=\columnwidth]{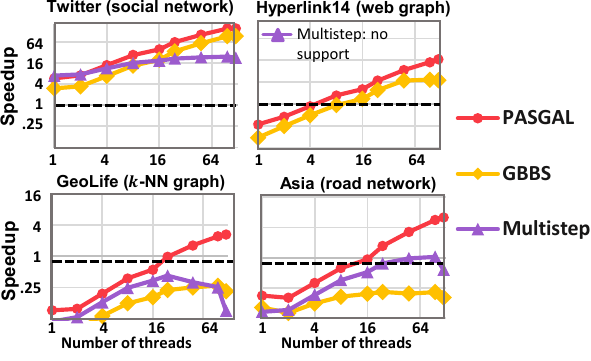}
  \caption{\small Speedup on \#processors for strongly connected component (SCC) algorithms over Tarjan's algorithm~\cite{tarjan1972depth} (sequential, always 1).  
  \normalfont 
  The tested algorithms include \pasgal{} (this paper), GBBS~\cite{gbbs2021}, and Multistep~\cite{slota2014bfs}. 
  We present four graphs of different types. The top two graphs have small diameters and the bottom two have larger diameters.
  %The largest graph Hyperlink14 has 1.7 billion vertices. 
   \label{fig:scc}}\label{fig:intro}
   \hide {GBBS~\cite{gbbs2021} 
  does not scale on some graphs due to theoretical inefficiency.  Tarjan-Vishkin~\cite{tarjan1985efficient} always scales (due to good span bound) but cannot run on large graphs (HL14~\cite{webgraph} has 1.7 billion vertices).  Due to theoretical efficiency, our algorithm~\cite{dong2023provably} is the fastest on all 27 tested graphs and always scales.
  }
\end{figure}

%Although the progress in hardware advance seems promising,
Despite the hardware advances, 
the increasing number of cores does not provide ``free'' performance improvement. 
We observed that many existing parallel systems suffer from scalability issues, both to more cores and to larger/more diverse graphs,
even in fundamental tasks such as breadth-first search (BFS), strongly connected components (SCC), biconnected components (BCC), and
single source shortest paths (SSSP).
We show an example of SCC algorithms in \cref{fig:intro}.
Tested on a 96-core machine,
existing systems~\cite{gbbs2021,slota2014bfs} scale well with fewer than 16 threads 
and/or on the well-studied power-law graphs with small diameters. 
Indeed, many of them focus on optimizing the performance of low-diameter graphs such as social networks. 
However, their performance can stop increasing (or even drop) with more threads, especially on large-diameter graphs.
On many graphs, they can perform worse than the sequential Tarjan's algorithm~\cite{tarjan1972depth}. 
Similar issues on other graph problems can be observed in \cref{fig:results}. 

One major reason for such performance degeneration is 
the high overhead of managing and synchronizing threads. 
While enabling the potential of better parallelism, more available cores also bring up great challenges and overhead.
Namely, \emph{parallelism comes at a cost}.
This is more pronounced when
using BFS-like primitives on large-diameter graphs:
when traversing the graph in BFS order, the number of rounds  (and thus the cost of scheduling and synchronizing threads between them)
is proportional to the diameter of the graph. 
As a result, when the diameter is large, the synchronization overhead can be more expensive than the computation.

\hide{
We identified two major reasons for such performance degeneration: 
high overhead of managing and synchronizing threads, 
and/or space inefficiency.
While enabling the potential of better parallelism, more available cores also bring up great challenges and synchronization overhead.
Namely, \emph{parallelism comes at a cost}.
This is more pronounced on
using BFS-like primitives on large-diameter graphs:
when traversing the graph in BFS order, the number of rounds (and thus the synchronization cost) 
is proportional to graph diameter. 
Meanwhile, some existing systems do not scale to large data not because of time, but space.  
%Although multicore parallelism avoids the network latencies as in distributed systems and cloud computing, it is highly limited by its valuable main memory size.
Memory is the most valuable resource for in-memory graph processing systems. 
Hence, space-inefficient algorithms may only process graphs with small to medium size, making the improvement from parallelism much narrower.
This can also be seen by \cref{fig:intro}: \gbbs{} incurs high synchronization cost due to the use of BFS, 
leading to unsatisfactory performance on large-diameter graphs. 
Tarjan-Vishkin avoids using BFS by introducing high space overhead, and does not scale to billion-scale graphs. }

We propose a new open-source library \emp{\pasgal}: the \emp{Parallel And Scalable Graph Algorithm Library}, that implements a list of graph algorithms that are scalable to diverse graph types, many processors, and large graphs. 
%These algorithms are for several problems where existing parallel solutions suffer severely from high synchronization costs, such as BFS, SCC, BCC, and SSSP.
\pasgal{} focuses on several problems where existing parallel solutions suffer severely from high synchronization costs, such as BFS, SCC, BCC, and SSSP.
%We provide a comprehensive evaluation of existing and \pasgal{'s} algorithms on a large variety of graphs.
We plan to include more algorithms in the future.

\hide{
To overcome the scalability issues, \pasgal{} employs three key techniques. 
The first is algorithm redesign to avoid using BFS. 
The second is a technique called \emph{vertical granularity control (VGC)} to hide scheduling overhead. 
%We further apply the idea to parallel BFS and achieved a new BFS implementation in \pasgal{}, which greatly improves the performance of BFS on large-diameter graphs. 
%Finally, some new parallel data structures are used to facilitate the algorithms. 
In \cref{sec:algo}, we introduce our techniques in more details. 
}

To overcome the scalability issues, the key technique in \pasgal{} 
is called \emph{vertical granularity control (VGC)} proposed in our recent paper~\cite{wang2023parallel}
to hide scheduling overhead. 
%We further apply the idea to parallel BFS and achieved a new BFS implementation in \pasgal{}, which greatly improves the performance of BFS on large-diameter graphs. 
Accordingly, we need to redesign algorithms and data structures
to facilitate VGC, as well as to synergistically optimize work, span and/or space usage. 
%Finally, some new parallel data structures are used to facilitate the algorithms. 
In \cref{sec:algo}, we introduce our techniques in more details.  

With VGC and other techniques in \cref{sec:algo}, 
\pasgal{} achieves high performance on a variety of graphs, especially large-diameter graphs.
%We test various types of graphs, including very large graphs with billions of vertices. 
For the aforementioned SCC problem, \pasgal{} exhibits good scalability (see \cref{fig:intro}),
and is faster than all previous parallel algorithms on all tested graphs (see \cref{fig:results}). 
%On SCC, BFS and SSSP, 
Overall, compared to the baselines, \pasgal{} is always competitive on small-diameter graphs, and is almost always the fastest on large diameter graphs. 
We discuss experimental results in \cref{sec:exp}. 
Our code is publicly available~\cite{pasgalcode}.
Full experimental results and more references are given in the Appendix. 
\myparagraph{Preliminaries.} Given a graph $G=(V,E)$,
we denote $n=|V|$ and $m=|E|$.
%We use $\nn(v)$ to denote the set of (out-)neighbors of $v$. 
%The hop distance between two vertices is the smallest number of edges on a path connecting them. 
%We use $D$ to denote the diameter of $G$, which is the largest hop distance among all pairs of vertices. 
%and $x$--$y$ be an edge between $x$ and $y$.
We use $D$ to denote the diameter of $G$. 
%A \defn{connected component (CC)} is a maximal subset in $V$
%such that every two vertices in it are connected by a path.
\hide{A \defn{breadth-first search (BFS)} visits all vertices starting from a source $s\in V$ based on their hop distances. 
A \defn{biconnected component (BCC)} is a maximal subset $C\subseteq V$
such that $C$ is connected and remains connected after removing any vertex $v\in C$.
A \defn{strongly connected component (SCC)} on a directed graph is a maximal subset $C\subseteq V$
such that for each $u,v\in C$, there exist a paths from $u$ to $v$, and a path from $v$ to $u$. }

Most algorithms in \pasgal{} are \emp{frontier-based} (see \cref{alg:BFS}). 
At a high level, the algorithm maintains a \emph{frontier}, which is a subset of vertices to be explored in each round. 
In round $i$, the algorithm \emph{processes} (visits their neighbors) the current frontier $\ff_i$ in parallel, 
and puts a subset of their neighbors to the next frontier $\ff_{i+1}$, determined by a certain condition. 
%For example, in BFS, if multiple vertices in $\ff_i$ attempt to add the same vertex to $\ff_{i+1}$, 
%a \cas{} is used to guarantee that only one will \emph{succeed}.
For example, in BFS, a vertex~$u$ will add its neighbor $v$ to the next frontier iff.\ $v$ has not been added to frontiers before, 
and $u$ is the first to add $v$ (based on some linearization order) in round $i$. 
The algorithm requires $\Omega(D)$ rounds. 

As mentioned, one key challenge of using parallel BFS or similar approaches is
the large cost to create and synchronize threads between rounds,
which is especially costly for large-diameter graphs (more rounds needed).
In this paper, we will show how \pasgal{} reduces
the scheduling overhead to achieve better parallelism.

\begin{algorithm}[t]
\small
\caption{Parallel Frontier-based Graph Algorithms\label{alg:BFS}}
\KwIn{A directed graph $G=(V,E)$ and an initial frontier $\ff_0$}
\SetKwFor{parForEach}{parallel\_for\_each}{do}{endfor}
\SetKwInOut{Maintains}{Maintains}
\DontPrintSemicolon
$i\gets 0$\\
\While {$\ff_{i}\ne \emptyset$}{
\parForEach{ $v\in \ff_i$}{
  %visit Put all unvisited neighbors of $v$ to $\ff_{i+1}$
  %\parForEach{$u\in \nn(v)$} {
  %  If a certain condition is true, put $u$ to $\ff_{i+1}$
  %}
  %Visit each of $v$'s neighbor $u$, put $u$ to $\ff_{i+1}$ if a certain condition is true.
  Visit all neighbors of $v$, put a subset of them to $\ff_{i+1}$.
}
%For each $v\in \ff_i$, put all unvisited neighbors (but avoid duplicates) of $v$ to $\ff_{i+1}$ in parallel\\
%Process all $v\in \ff_i$ and their edges in parallel, put all their unvisited neighbors (but avoid duplicates) to $\ff_{i+1}$\\
$i\gets i+1$
}
\end{algorithm}

\section{Algorithms} 
\label{sec:algo} 

We now introduce the algorithms in \pasgal{}.
For page limit, we only elaborate on SCC and briefly overview the others. 

\subsection{Parallel SCC}
\label{sec:scc}

Most existing parallel SCC algorithms are based on \emp{reachability search},
which finds all vertices $u$ that are reachable from a given vertex $v$. 
In most existing implementations, reachability searches are performed by
BFS from $v$, which requires $O(D)$ rounds to finish. 
This directly implies several (interrelated) challenges on large-diameter graphs. 
First, 
this incurs many rounds of distributing and synchronizing threads with high overhead. 
Second, many real-world large-diameter graphs (e.g., road networks) are sparse 
with small average degrees.
%indicating a small neighborhood size for each vertex.
As a result, every parallel task (processing one vertex in the frontier) is small, 
and the cost of scheduling the thread may dominate the actual computation.
Finally, because of sparsity, each frontier size is also likely small, 
which makes the algorithm unable to utilize all threads in the hardware. 

\myparagraph{Algorithm Redesign.} To resolve this challenge, \pasgal{} uses a recent SCC algorithm \cite{wang2023parallel}. 
The idea is to observe that a reachability search does not require a strong BFS order.
Therefore, one can relax the BFS order and visit vertices in an arbitrary order. 
In this way, the algorithm employs a technique called \emp{vertical granularity control}
to hide scheduling overhead, as introduced below. 

\myparagraph{Vertical Granularity Control.} 
\emph{Granularity control} (a.k.a.\ coarsening) is widely used in parallel programming,
% Audiences familiar with parallel programming must know the concept of \emph{granularity control} (a.k.a.\ coarsening), 
which also aims to avoid the overhead caused by generating unnecessary parallel tasks.
For computations with sufficient parallelism, e.g., a parallel for-loop or divide-and-conquer algorithm of size $n\gg P$ where $P$ is the number of processors, 
a common practice is to stop recursively creating parallel tasks at a certain subproblem size and switch to a sequential execution to hide the scheduling overhead.

Inspired by this, the idea of VGC is also to increase each task size to hide the scheduling overhead. 
%but we merge the computation across \emph{different levels} to acquire more work and saturate all processors in each round. 
%In this way, we break the synchronization points and reduce scheduling overhead. 
For reachability searches, we simply perform a \emph{local search} to visit at least $\tau$ vertices, possibly in multiple hops. 
While globally the vertices are not visited in the BFS order, this does not affect the correctness of reachability. 
Note that here $\tau$ is equivalent to the base-case task size of granularity control, and is a tunable parameter.
In this way, VGC 1) greatly reduces the number of rounds, since each round may advance multiple hops from the current frontier, and 2) quickly accumulates a large frontier size since the next frontier contains multiple-hop neighbors from the current frontier, and thus yields sufficient parallel tasks throughout the algorithm.
Both outcomes effectively reduce (or hide) synchronization costs and allow for much better parallelism. 

\myparagraph{Data Structure Design.} Another useful technique for the SCC algorithm is a data structure called \emph{hash bag}~\cite{wang2023parallel}. 
It maintains a dynamic set of vertices as the frontiers for parallel graph algorithms. 
For page limit, we refer the readers to \cite{wang2023parallel} for more details. 

\begin{figure*}[!h!t]
  \small
  \centering
  \includegraphics[width=0.9\textwidth]{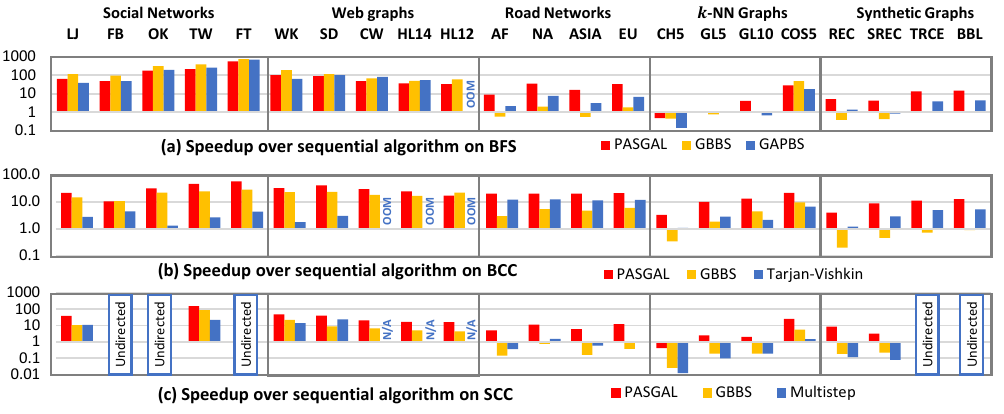}
  \caption{Speedup of parallel algorithms over the standard sequential algorithm. $y$-axis is in log-scale. Bars below 1.0 mean the parallel algorithm is slower than a sequential one. 
  Some bars are invisible because they are close to 1. ``N/A'': not applicable. ``OOM'': out-of-memory.
  }
  \label{fig:results}
\end{figure*}

\begin{table}[t]
  \centering
  \small
    %\begin{tabular}{@{ }l@{ }|@{ }r@{ }r@{ }r@{ }r@{ }|@{ }r@{ }r@{ }r@{ }r@{ }r@{ }}
    \begin{tabular}{@{}l@{ }|@{ }*4{R{}}|@{ }*5{R{0.6cm}}@{ }}
    \toprule
          && \multicolumn{2}{c}{\textbf{Road}}& & &\multicolumn{3}{c}{\textbf{Social}} &\\
          & \textbf{AF} & \textbf{NA} & \textbf{AS} & \textbf{EU} & \textbf{LJ} & \textbf{\underline{FB}} & \textbf{\underline{OK}} & \textbf{TW} & \textbf{\underline{FS}} \\
    \midrule
    {$n$} & 33.5M & 87.0M & 95.7M & 131M  & 4.85M & 59.2M & 3.07M & 41.7M & 65.6M \\
    {$m'$} & 44.8M & 113M  & 123M  & 169M  & 69.0M & N/A   & N/A   & 1.47B & 3.61B \\
    {$m$} & 88.9M & 220M  & 244M  & 333M  & 85.7M & 185M  & 234M  & 2.41B & 3.61B \\
    {$D'$} & 8276 & 9337 & 16660 & 11814 & 22 & N/A & N/A & 18 & N/A \\
    {$D$} & 3948 & 4835 & 8794 & 4410 & 19 & 22 & 9 & 22 & 37 \\
    \midrule
     && \multicolumn{2}{c}{\textbf{$k$-NN}}& && \multicolumn{3}{c}{\textbf{Web}} &\\
     & \textbf{CH5} & \textbf{GL5} & \textbf{GL10} & \textbf{COS5} & \textbf{WK} & \textbf{SD} & \textbf{CW} & \textbf{HL14} & \textbf{HL12} \\
     \midrule
    {$n$} & 4.21M & 24.9M & 24.9M & 321M  & 6.63M & 89.2M & 978M  & 1.72B & 3.56B \\
    {$m'$} & 21.0M & 124M  & 249M  & 1.61B & 165M  & 2.04B & 42.4B & 64.4B & 129B \\
    {$m$} & 29.7M & 157M  & 310M  & 1.96B & 300M  & 3.88B & 74.7B & 124B  & 226B \\
    {$D'$} & 5683 & 17268 & 13982 & 1390 & 62 & 145 & 506 & 800 & 5279 \\
    {$D$} & 14479 & 21601 & 9053 & 1180 & 9 & 35 & 254 & 366 & 650 \\
    \midrule
     && \multicolumn{2}{c}{\textbf{Synthetic}} & &\multicolumn{5}{l}{Underline: undirected graphs} \\
     & \textbf{REC} & \textbf{SREC} & \textbf{\underline{TRCE}} & \textbf{\underline{BBL}} & \multicolumn{5}{l}{$m'$: \#edges in directed graphs} \\
     \cmidrule{1-5}
    {$n$} & 100M  & 100M  & 16.0M & 21.2M & \multicolumn{5}{l}{$m$: \#edges in undirected or} \\
    {$m'$} & 297M  & 204M  & N/A & N/A   & \multicolumn{5}{l}{\quad symmetrized graph} \\
    {$m$} & 400M  & 336M  & 48.0M & 63.6M & \multicolumn{5}{l}{$D'$: diameter of the directed graph} \\
    {$D'$} & 59075 & 102151 & N/A & N/A   &  \multicolumn{5}{l}{$D$: diameter of the undirected or} \\
    {$D$} & 50500 & 54843 & 5527 & 7849 &   \multicolumn{5}{l}{\quad symmetrized graph} \\
    \bottomrule
    \end{tabular}%
      \caption{Tested graphs. Since it is hard to obtain the exact value of $D$ and $D'$, 
      the number shown is a lower bound obtained by at least 1000 sampled searches on each graph.}
  \label{tab:graph-stats}%
\end{table}% 

\subsection{Other Algorithms}\label{sec:bcc}

\myparagraph{Parallel SSSP.} The SSSP algorithm in \pasgal{} is based on the \emph{stepping algorithm framework}~\cite{dong2021efficient}, and uses VGC and hash bags introduced in \cref{sec:scc} to accelerate the frontier traversing. 
%When visiting a vertex $v$ in the frontier, we attempt to visit and relax vertices within multiple hops from $v$. 
%When the relaxation succeeds, the newly visited vertex will be added to the next frontier. 

\myparagraph{Parallel BFS.} Using VGC, we implemented a new BFS algorithm in \pasgal{}. 
We also use hash bags to maintain the frontiers. 
%Note that different from SCC, a BFS algorithm itself requires to maintain strong BFS order when visiting vertices.
%To do this with local search, we use a distance array to maintain the hop distance for each vertex from the source
Our BFS algorithm is similar to SSSP where the output distance is the hop distance from the source. 
For any vertex encountered in a local search from vertex $v$, we add it to the corresponding frontier if its hop distance can be updated by $v$. 
Note that due to local search, a vertex can be visited multiple times instead of exactly once as in standard BFS.
This is because the distance of a vertex $v$ updated by a local search is not necessarily the shortest distance of $v$,
leading to extra work. 
To reduce this overhead, one special technique for BFS is that we maintain multiple frontiers, where frontier $i$ maintains
vertices with distance $2^i$ from the current frontier. 
%We combine and split the frontiers when necessary. 
In this way, we obtain the benefit of BFS by exploring multiple hops in one round,
and also avoid visiting too many ``unready'' vertices in the frontier. 
We also use the direction optimization~\cite{Beamer12} to improve performance. 

\myparagraph{Parallel Biconnectivity.} Different from other problems, 
the major performance gain of the BCC algorithm in \pasgal{} is due to algorithm redesign
to achieve \emph{stronger theoretical bounds}. 
The performance bottleneck of previous BCC algorithms either comes from 
the use of BFS that requires $O(D)$ rounds of global synchronizations (e.g., \gbbs{}~\cite{gbbs2021}),
or requires $O(m)$ auxiliary space and does not scale to large graphs (e.g., Tarjan-Vishkin~\cite{tarjan1985efficient}). 
\pasgal{} uses the FAST-BCC algorithm in~\cite{dong2023provably}. 
By redesigning the algorithm, FAST-BCC avoids the use of BFS, and achieves $O(n+m)$ work, polylogarithmic span, and $O(n)$ auxiliary space. 
We also use VGC and hash bags to further improve the performance. 

\section{Experimental Results}
\label{sec:exp} 

\myparagraph{Library Design.}
We release the code of \pasgal~\cite{pasgalcode}. 
\pasgal{} is implemented in C++ using ParlayLib~\cite{blelloch2020parlaylib} for fork-join parallelism and some parallel primitives (e.g., sorting). 
The four algorithms (BFS, BCC, SCC and SSSP) are provided in the subdirectories. 
A readme file about compiling and running the library is provided in the repository. 
\pasgal{} supports two different graph formats: the binary format (\texttt{.bin}) from \gbbs{}~\cite{gbbs2021}, and the adjacency graph format (\texttt{.adj}) from the Problem-Based Benchmark Suite (PBBS)~\cite{anderson2022problem}. 

\myparagraph{Setup.}
We run our experiments on a 96-core (192 hyperthreads) machine with four Intel Xeon Gold 6252 CPUs and 1.5 TB of main memory.
We use \texttt{numactl -i all} in parallel experiments.
%We run each test for 5 times and report the median.

We tested on 22 graphs, including social networks, web graphs, road graphs, \knn{} graphs, and synthetic graphs.
All the graphs are from existing research papers and public datasets. %~\cite{}. 
The graph information is given in \cref{alg:BFS}.
For page limit, we provide the full graph information and corresponding citations in the Appendix. 
%We call each test on a graph-implementation combination an \emph{instance}.
We call the social and web graphs \emph{\lowdiam{} graphs} as they have diameters mostly within a few hundred. We call the road, \knn{}, and synthetic graphs \emph{\largediam{} graphs} as their diameters are mostly more than a thousand. 
%When comparing the \emph{average} running times across multiple graphs, we always take the \defn{geometric mean} of the numbers. 
We symmetrize directed graphs for testing BCC.  
SCC does not apply to undirected graphs. 

We present the performance comparison in \cref{fig:results}. 
For page limit, we only show results for SCC, BCC, and BFS. 
For each problem, we also implement the standard sequential algorithm as the baseline, which is a queue-based solution for BFS, Tarjan's algorithm for SCC~\cite{tarjan1972depth}, 
and the Hopcroft-Tarjan algorithm for BCC~\cite{hopcroft1973algorithm}. 
We compute the relative speedup for each parallel implementation over the sequential algorithm, and present them in \cref{fig:results}. 
The $y$-axis is in log-scale. Bars below 1 mean that the parallel algorithm is \emph{slower than the sequential implementation}. 
The baselines include state-of-the-art graph libraries and implementations, including \gbbs{}~\cite{gbbs2021}, \gapbs{}~\cite{beamer2015gap}, 
Tarjan-Vishkin from~\cite{dong2023provably}, and Multistep~\cite{slota2014bfs}. 

For all tested problems and all graphs, \pasgal{} is always competitive on small-diameter graphs: 
across all graphs, \pasgal{} is within $1.3\times$ of the running time compared to the fastest baseline on BCC, 
$2\times$ on BFS, and always faster than all baselines on SCC. 
Parallel BFS on social networks is one of the most well-studied parallel graph algorithms, 
and all parallel algorithms achieve superlinear speedup on some social networks due to various optimizations (e.g., the direction optimization~\cite{Beamer12}). 
\pasgal{} achieves good scalability and is 49-570$\times$ faster than the standard sequential algorithm using 192 threads. 
An interesting future direction is to further make the BFS performance of \pasgal{} match the best parallel implementation. 

On large-diameter graphs, \pasgal{} achieves much better performance than \emph{all baselines}.
On BCC, due to theoretical efficiency, \pasgal{} consistently outperforms the sequential Hopcroft-Tarjan algorithm on all graphs.
It is up to 3.45$\times$ faster than the best baseline on each graph. 
On SCC and BFS, \pasgal{} is always faster than the sequential baseline except for one graph CH5,
which has very large diameter compared to its small size. 
Different from BCC, 
our SCC and BFS algorithms do not have strong span bound. 
Using VGC can only alleviate the scalability issue on large-diameter graphs,
but may still be unable to eliminate the issue on adversarial graphs (e.g., a chain). 
Still, on most real-world large-diameter graphs, \pasgal{} is almost always the fastest among all parallel implementations, and is up to $5\times$
faster than the best baseline on BFS, and up to 46$\times$ on SCC. 
%and can be much faster on large-diameter graphs (up to $5\times$ faster than the fastest baseline on BFS, $3.2\times$ on BCC and xx on SCC). 

\hide{
For all tested problems and all graphs, \pasgal{} is always competitive on small-diameter graphs,
and can be much faster on large-diameter graphs.
Compared to \gbbs{} that is used }

\section{Conclusion and Future Work} 

In this paper, we present \pasgal{}, a scalable graph library specially designed to improve performance on large-diameter graphs.
As mentioned, some interesting future directions include further seeking new ideas to improve the performance of BFS on small-diameter graphs
that also work well with VGC, as well as further improving the performance for BFS and SCC on very large-diameter graphs. 
In addition, we believe the techniques in current \pasgal{} can be extended to more problems, including $k$-core and other peeling algorithms, 
$k$-connectivity, point-to-point shortest paths, etc. We plan to add them to \pasgal{} in the future. 

\section*{Acknowledgement}
This work is supported by NSF grants CCF-2103483, CCF-2238358, CCF-2339310, and IIS-2227669, the UCR Regents Faculty Development Award, and the Google Research Scholar Program.

\small

%\bibliographystyle{ACM-Reference-Format}
%\bibliography{bib/strings,bib/main}

\appendix
\section{More Experimental Results}
We present the original running time in our experiments and more detailed graph information here.
% We will put this document online for reference if this brief announcement is accepted. 

\cref{tab:graphfull} presents the information of all graphs tested in the paper. 
\cref{tab:bcctime,tab:bfstime,tab:scctime} presents
the running time of all tested algorithms on biconnected components (BCC),
breadth-first search (BFS), and strongly connected components (SCC), respectively. 

For each table, the last column is always the standard \emph{sequential} algorithm (noted with ``*'').
In the figure in the main paper, the relative speedup are based on this sequential algorithm. 
For all running times, lower is better. ``o.o.m'' means out-of-memory. ``n.s.'' means no-support. 

% Table generated by Excel2LaTeX from sheet 'graph-full'
\begin{table*}[htbp]
  \centering
\small
    \begin{tabular}{rrrrrrrr}
    \toprule
          &       & \multicolumn{1}{c}{$\boldsymbol{n}$} & \multicolumn{1}{c}{$\boldsymbol{m'}$} & \multicolumn{1}{c}{$\boldsymbol{m}$} & \multicolumn{1}{c}{$\boldsymbol{D'}$} & \multicolumn{1}{c}{$\boldsymbol{D}$} & \multicolumn{1}{c}{\textbf{Notes}} \\
    \midrule
    \multicolumn{1}{c}{\multirow{5}[2]{*}{\begin{sideways}\textbf{Social}\end{sideways}}} & \multicolumn{1}{l}{\textbf{LJ}} & 4,847,571 & 68,993,773 & 85,702,474 & 22    & 19    & \multicolumn{1}{l}{soc-LiveJournal1~\cite{backstrom2006group}} \\
    \multicolumn{1}{c}{} & \multicolumn{1}{l}{\textbf{\underline{FB}}} & 59,216,214 & N/A   & 185,044,024 & N/A   & 22    & \multicolumn{1}{l}{socfb-konect~\cite{nr,traud2012social,red2011comparing}} \\
    \multicolumn{1}{c}{} & \multicolumn{1}{l}{\textbf{\underline{OK}}} & 3,072,627 & N/A   & 234,370,166 & N/A   & 9     & \multicolumn{1}{l}{com-orkut~\cite{yang2015defining}} \\
    \multicolumn{1}{c}{} & \multicolumn{1}{l}{\textbf{TW}} & 41,652,231 & 1,468,365,182 & 2,405,026,092 & 18    & 22    & \multicolumn{1}{l}{Twitter~\cite{kwak2010twitter}} \\
    \multicolumn{1}{c}{} & \multicolumn{1}{l}{\textbf{\underline{FS}}} & 65,608,366 & N/A   & 3,612,134,270 & N/A   & 37    & \multicolumn{1}{l}{Friendster~\cite{yang2015defining}} \\
    \midrule 
    \multicolumn{1}{c}{\multirow{5}[2]{*}{\begin{sideways}\textbf{Web}\end{sideways}}} & \multicolumn{1}{l}{\textbf{WK}} & 6,625,370 & 165,206,104 & 300,331,770 & 62    & 9     & \multicolumn{1}{l}{enwiki-2023~\cite{BoVWFI,Boldi-2011-layered}} \\
    \multicolumn{1}{c}{} & \multicolumn{1}{l}{\textbf{SD}} & 89,247,739 & 2,043,203,933 & 3,880,015,728 & 145   & 35    & \multicolumn{1}{l}{sd-arc~\cite{webgraph}} \\
    \multicolumn{1}{c}{} & \multicolumn{1}{l}{\textbf{CW}} & 978,408,098 & 42,574,107,469 & 74,744,358,622 & 506   & 254   & \multicolumn{1}{l}{ClueWeb~\cite{webgraph}} \\
    \multicolumn{1}{c}{} & \multicolumn{1}{l}{\textbf{HL14}} & 1,724,573,718 & 64,422,807,961 & 124,141,874,032 & 800   & 366   & \multicolumn{1}{l}{Hyperlink14~\cite{webgraph}} \\
    \multicolumn{1}{c}{} & \multicolumn{1}{l}{\textbf{HL12}} & 3,563,602,789 & 128,736,914,167 & 225,840,663,232 & 5279  & 650   & \multicolumn{1}{l}{Hyperlink12~\cite{webgraph}} \\
    \midrule 
    \multicolumn{1}{c}{\multirow{4}[2]{*}{\begin{sideways}\textbf{Road}\end{sideways}}} & \multicolumn{1}{l}{\textbf{AF}} & 33,493,259 & 44,773,338 & 88,929,770 & 8276  & 3948  & \multicolumn{1}{l}{Open Street Map (OSM) Africa~\cite{roadgraph}} \\
    \multicolumn{1}{c}{} & \multicolumn{1}{l}{\textbf{NA}} & 86,951,513 & 112,869,708 & 220,285,922 & 9337  & 4835  & \multicolumn{1}{l}{Open Street Map (OSM) North America~\cite{roadgraph}} \\
    \multicolumn{1}{c}{} & \multicolumn{1}{l}{\textbf{AS}} & 95,735,401 & 122,836,037 & 243,624,688 & 16660 & 8794  & \multicolumn{1}{l}{Open Street Map (OSM) Asia~\cite{roadgraph}} \\
    \multicolumn{1}{c}{} & \multicolumn{1}{l}{\textbf{EU}} & 130,655,972 & 168,541,220 & 332,587,928 & 11814 & 4410  & \multicolumn{1}{l}{Open Street Map (OSM) Europe~\cite{roadgraph}} \\
    \midrule 
    \multicolumn{1}{c}{\multirow{4}[2]{*}{\begin{sideways}\boldmath{}\textbf{$k$NN}\unboldmath{}\end{sideways}}} & \multicolumn{1}{l}{\textbf{CH5}} & 4,208,261 & 21,041,305 & 29,650,038 & 5683  & 14479 & 
    \multicolumn{1}{l}{Chem~\cite{fonollosa2015reservoir,wang2021geograph}, $k=5$} \\
    \multicolumn{1}{c}{} & \multicolumn{1}{l}{\textbf{GL5}} & 24,876,978 & 124,384,890 & 157,442,032 & 17268 & 21601 & \multicolumn{1}{l}{GeoLife~\cite{geolife,wang2021geograph}, $k=5$} \\
    \multicolumn{1}{c}{} & \multicolumn{1}{l}{\textbf{GL10}} & 24,876,978 & 248,769,780 & 309,743,322 & 13982 & 9053  & \multicolumn{1}{l}{GeoLife~\cite{geolife,wang2021geograph}, $k=10$} \\
    \multicolumn{1}{c}{} & \multicolumn{1}{l}{\textbf{COS5}} & 321,065,547 & 1,605,327,735 & 1,957,750,718 &  1390   & 1180  & \multicolumn{1}{l}{Cosmo50~\cite{cosmo50,wang2021geograph}, $k=5$} \\
    \midrule 
    \multicolumn{1}{c}{\multirow{4}[2]{*}{\begin{sideways}\textbf{Synthetic}\end{sideways}}} & \multicolumn{1}{l}{\textbf{REC}} & 100,000,000 & 297,418,030 & 400,000,000 & 59075   & 50500 & \multicolumn{1}{l}{$10^3\times 10^5$ Grid~\cite{wang2023parallel}} \\
    \multicolumn{1}{c}{} & \multicolumn{1}{l}{\textbf{SREC}} & 100,000,000 & 203,785,880 & 335,787,820 & 102151  & 54843 & \multicolumn{1}{l}{Sampled REC~\cite{wang2023parallel}} \\
    \multicolumn{1}{c}{} & \multicolumn{1}{l}{\textbf{\underline{TRCE}}} & 16,002,413 & N/A   & 47,997,626 & N/A   & 5527  & \multicolumn{1}{l}{Huge traces~\cite{nr}} \\
    \multicolumn{1}{c}{} & \multicolumn{1}{l}{\textbf{\underline{BBL}}} & 21,198,119 & N/A   & 63,580,358 & N/A   & 7849  & \multicolumn{1}{l}{Huge bubbles~\cite{nr}} \\
    \bottomrule
    \end{tabular}%
    \caption{\textbf{Information of all tested graphs.} 
    $n$: \#vertices.
    $m$: \#edges in the undirected or symmetrized graph. $m'$: \# directed edges.
    $D$: diameter of the undirected or symmetrized graph. 
    $D'$: diameter of the directed graph. 
    Undirected graphs are underlined.
    \label{tab:graphfull}
    }
\end{table*}%
\clearpage

% Table generated by Excel2LaTeX from sheet 'bcc'
\begin{table*}[htbp]
  \centering
    \begin{tabular}{rrrrrr}
    \toprule
    \textbf{} & \multicolumn{1}{c}{\textbf{Graph}} & \multicolumn{1}{c}{\textbf{PASGAL}} & \multicolumn{1}{c}{\textbf{GBBS}} & \multicolumn{1}{c}{\textbf{Tarjan-Vishkin}} & \multicolumn{1}{c}{\textbf{Hopcroft-Tarjan*}} \\
    \midrule
    \multicolumn{1}{c}{\multirow{5}[2]{*}{\begin{sideways}\textbf{Social}\end{sideways}}} & \textbf{LJ} & 0.112 & 0.165 & 0.857 & 2.38 \\
    \multicolumn{1}{c}{} & \textbf{FB} & 0.835 & 0.825 & 1.95  & 8.68 \\
    \multicolumn{1}{c}{} & \textbf{OK} & 0.106 & 0.15  & 2.46  & 3.3 \\
    \multicolumn{1}{c}{} & \textbf{TW} & 1.44  & 2.72  & 24.6  & 65.6 \\
    \multicolumn{1}{c}{} & \textbf{FT} & 3.17  & 6.26  & 40.7  & 177.6 \\
    \midrule
    \multicolumn{1}{c}{\multirow{5}[2]{*}{\begin{sideways}\textbf{Web}\end{sideways}}} & \textbf{WK} & 0.175 & 0.248 & 3.1   & 5.58 \\
    \multicolumn{1}{c}{} & \textbf{SD} & 3.16  & 5.47  & 41.7  & 127.9 \\
    \multicolumn{1}{c}{} & \textbf{CW} & 23.8  & 39.4  & o.o.m.   & 704.5 \\
    \multicolumn{1}{c}{} & \textbf{HL14} & 33.7  & 49.3  & o.o.m.   & 818.3 \\
    \multicolumn{1}{c}{} & \textbf{HL12} & 129.8 & 102.4 & o.o.m.   & 2216.6 \\
    \midrule
    \multicolumn{1}{c}{\multirow{4}[2]{*}{\begin{sideways}\textbf{Road}\end{sideways}}} & \textbf{AF} & 0.859 & 5.77  & 1.42  & 17.1 \\
    \multicolumn{1}{c}{} & \textbf{NA} & 2.19  & 8.19  & 3.62  & 44.2 \\
    \multicolumn{1}{c}{} & \textbf{ASIA} & 2.46  & 10.4  & 4.33  & 49.1 \\
    \multicolumn{1}{c}{} & \textbf{EU} & 3.32  & 11.4  & 5.84  & 69 \\
    \midrule
    \multicolumn{1}{c}{\multirow{4}[2]{*}{\begin{sideways}\boldmath{}\textbf{$k$NN}\unboldmath{}\end{sideways}}} & \textbf{CH5} & 0.141 & 1.31  & 0.418 & 0.465 \\
    \multicolumn{1}{c}{} & \textbf{GL5} & 0.501 & 2.65  & 1.73  & 4.97 \\
    \multicolumn{1}{c}{} & \textbf{GL10} & 0.558 & 1.66  & 3.37  & 7.36 \\
    \multicolumn{1}{c}{} & \textbf{COS5} & 8.28  & 18.4  & 26.5  & 175.8 \\
    \midrule
    \multicolumn{1}{c}{\multirow{4}[2]{*}{\begin{sideways}\textbf{Synthetic}\end{sideways}}} & \textbf{REC} & 1.47  & 28.7  & 4.81  & 5.95 \\
    \multicolumn{1}{c}{} & \textbf{SREC} & 1.44  & 26.3  & 4.32  & 12.5 \\
    \multicolumn{1}{c}{} & \textbf{TRCE} & 0.302 & 4.52  & 0.674 & 3.36 \\
    \multicolumn{1}{c}{} & \textbf{BBL} & 0.406 & 5.62  & 0.958 & 5.1 \\
    %\bottomrule
    %\toprule
    \midrule
    \multicolumn{6}{l}{\bf Geometric Mean:}\\
        %\multirow{5}{*}{\begin{sideways}\bf Geomean\end{sideways}} 
    \midrule
        & \bf Social & 0.63  &  0.913  & 5.50 & 9.53 \\
          & \bf Web   & 4.97  & 7.06 & -  & 112 \\
          & \bf Road  & 1.98  & 8.64  & 3.38  & 20.7 \\
          & \bf \boldmath $k$NN \unboldmath  & 0.756 & 3.21  & 2.84  & 4.55 \\
          & \bf Synthetic & 0.714 & 11.8  & 1.91  & 3.68 \\
    \bottomrule
    \end{tabular}%
      \caption{Running time of all tested algorithms on biconnected components (BCC). 
      The parallel implementations include PASGAL (this paper), GBBS~\cite{gbbs2021}, and Tarjan-Vishkin~\cite{tarjan1985efficient} (implementation from~\cite{dong2023provably}).
      The sequential baseline is the Hopcroft-Tarjan algorithm~\cite{hopcroft1973algorithm}. 
      }
  \label{tab:bcctime}%
\end{table*}% 
\clearpage

% Table generated by Excel2LaTeX from sheet 'scc'
\begin{table}[htbp]
  \centering
    \begin{tabular}{cr|rrrr}
    \toprule
    \textbf{} & \multicolumn{1}{c}{\textbf{Graph}} & \multicolumn{1}{c}{\textbf{PASGAL}} & \multicolumn{1}{c}{\textbf{GBBS}} & \multicolumn{1}{c}{\textbf{Multistep}} & \multicolumn{1}{c}{\textbf{Tarjan*}} \\
    \midrule
    \multirow{5}[2]{*}{\begin{sideways}\textbf{Social}\end{sideways}} & \multicolumn{1}{l}{\textbf{LJ}} & 0.033 & 0.122 & 0.117 & 1.31 \\
          & \multicolumn{1}{l}{\textbf{FB}} & \multicolumn{4}{c}{Undirected Graph} \\
          & \multicolumn{1}{l}{\textbf{OK}} & \multicolumn{4}{c}{Undirected Graph} \\
          & \multicolumn{1}{l}{\textbf{TW}} & 0.192 & 0.332 & 1.31  & 30.2 \\
          & \multicolumn{1}{l}{\textbf{FT}} & \multicolumn{4}{c}{Undirected Graph} \\
    \midrule      
    \multirow{5}[2]{*}{\begin{sideways}\textbf{Web}\end{sideways}} & \multicolumn{1}{l}{\textbf{WK}} & 0.054 & 0.120 & 0.185 & 2.71 \\
          & \multicolumn{1}{l}{\textbf{SD}} & 1.06  & 4.93  & 1.84  & 44.5 \\
          & \multicolumn{1}{l}{\textbf{CW}} & 12.6  & 38.8  & n.s.  & 267 \\
          & \multicolumn{1}{l}{\textbf{HL14}} & 18.8  & 63.6  & n.s.  & 326 \\
          & \multicolumn{1}{l}{\textbf{HL12}} & 96.8  & 364.0 & n.s.  & 1620 \\
    \midrule
    \multirow{4}[2]{*}{\begin{sideways}\textbf{Road}\end{sideways}} & \multicolumn{1}{l}{\textbf{AF}} & 1.40  & 50.3  & 19.9  & 7.32 \\
          & \multicolumn{1}{l}{\textbf{NA}} & 1.64  & 24.6  & 12.4  & 19.2 \\
          & \multicolumn{1}{l}{\textbf{ASIA}} & 3.34  & 129   & 34.1  & 20.8 \\
          & \multicolumn{1}{l}{\textbf{EU}} & 2.34  & 76.2  & 31.8  & 29.2 \\
    \midrule
    \multirow{4}[2]{*}{\begin{sideways}\boldmath{}\textbf{$k$NN}\unboldmath{}\end{sideways}} & \multicolumn{1}{l}{\textbf{CH5}} & 0.528 & 8.45  & 18.0  & 0.224 \\
          & \multicolumn{1}{l}{\textbf{GL5}} & 0.903 & 11.9  & 23.7  & 2.34 \\
          & \multicolumn{1}{l}{\textbf{GL10}} & 1.62  & 17.5  & 17.7  & 3.41 \\
          & \multicolumn{1}{l}{\textbf{COS5}} & 3.27  & 14.9  & 58.0  & 87.1 \\
    \midrule
    \multirow{4}[2]{*}{\begin{sideways}\textbf{Synthetic}\end{sideways}} & \multicolumn{1}{l}{\textbf{REC}} & 0.892 & 41.3  & 67.3  & 7.91 \\
          & \multicolumn{1}{l}{\textbf{SREC}} & 2.02  & 30.2  & 82.2  & 6.70 \\
          & \multicolumn{1}{l}{\textbf{TRCE}} & \multicolumn{4}{c}{Undirected Graph} \\
          & \multicolumn{1}{l}{\textbf{BBL}} & \multicolumn{4}{c}{Undirected Graph} \\
    \midrule
    \multicolumn{6}{c}{\textbf{Geometric Mean}} \\
    \midrule
    \multirow{5}[2]{*}{} & \multicolumn{1}{l}{\textbf{Social}} & 0.080 & 0.201 & 0.391 & 6.30 \\
          & \multicolumn{1}{l}{\textbf{Web}} & 4.20  & 14.0  & -     & 112 \\
          & \multicolumn{1}{l}{\textbf{Road}} & 2.06  & 59.0  & 22.8  & 17.1 \\
          & \multicolumn{1}{l}{\boldmath\textbf{$k$NN}\unboldmath} & 1.26  & 12.7  & 25.7  & 3.53 \\
          & \multicolumn{1}{l}{\textbf{Synthetic}} & 1.34  & 35.3  & 74.4  & 7.28 \\
    \bottomrule
    \end{tabular}%
      \caption{Running time of all tested algorithms on strongly connected components (SCC). 
      The parallel implementations include PASGAL (this paper), GBBS~\cite{gbbs2021}, and Multistep~\cite{slota2014bfs}.
      Multistep does not support the three largest graph because the number of vertices in CW, HL14, and HL12 are larger than 32-bit integers. 
      The sequential baseline is the Tarjan's algorithm~\cite{tarjan1972depth}. }
  \label{tab:scctime}%
\end{table}% 

% Table generated by Excel2LaTeX from sheet 'bfs'
\begin{table}[htbp]
  \centering
    \begin{tabular}{crrrrr}
    \toprule
    \textbf{} & \multicolumn{1}{c}{\textbf{Graph}} & \multicolumn{1}{c}{\textbf{PASGAL}} & \multicolumn{1}{c}{\textbf{GBBS}} & \multicolumn{1}{c}{\textbf{GAPBS}} & \multicolumn{1}{c}{\textbf{Queue-based*}} \\
    \midrule
    \multirow{5}[2]{*}{\begin{sideways}\textbf{Social}\end{sideways}} & \multicolumn{1}{l}{\textbf{LJ}} & 0.018 & 0.010 & 0.030 & 1.19 \\
          & \multicolumn{1}{l}{\textbf{FB}} & 0.104 & 0.052 & 0.104 & 5.14 \\
          & \multicolumn{1}{l}{\textbf{OK}} & 0.012 & 0.007 & 0.011 & 2.19 \\
          & \multicolumn{1}{l}{\textbf{TW}} & 0.136 & 0.077 & 0.116 & 30.2 \\
          & \multicolumn{1}{l}{\textbf{FT}} & 0.214 & 0.163 & 0.173 & 122 \\
    \midrule
    \multirow{5}[2]{*}{\begin{sideways}\textbf{Web}\end{sideways}} & \multicolumn{1}{l}{\textbf{WK}} & 0.026 & 0.014 & 0.041 & 2.67 \\
          & \multicolumn{1}{l}{\textbf{SD}} & 0.471 & 0.366 & 0.421 & 43.1 \\
          & \multicolumn{1}{l}{\textbf{CW}} & 5.69  & 4.10  & 3.44  & 283.3 \\
          & \multicolumn{1}{l}{\textbf{HL14}} & 4.19  & 3.12  & 2.83  & 158 \\
          & \multicolumn{1}{l}{\textbf{HL12}} & 22.5  & 12.2  & o.o.m.   & 759 \\
    \midrule
    \multirow{4}[2]{*}{\begin{sideways}\textbf{Road}\end{sideways}} & \multicolumn{1}{l}{\textbf{AF}} & 0.05  & 0.7   & 0.2   & 0.45 \\
          & \multicolumn{1}{l}{\textbf{NA}} & 0.31  & 5.4   & 1.4   & 11.0 \\
          & \multicolumn{1}{l}{\textbf{ASIA}} & 0.18  & 5     & 0.9   & 3.0 \\
          & \multicolumn{1}{l}{\textbf{EU}} & 0.49  & 8.8   & 2.3   & 16.3 \\
    \midrule
    \multirow{4}[2]{*}{\begin{sideways}\boldmath{}\textbf{$k$NN}\unboldmath{}\end{sideways}} & \multicolumn{1}{l}{\textbf{CH5}} & 0.095 & 0.10  & 0.3   & 0.047 \\
          & \multicolumn{1}{l}{\textbf{GL5}} & 0.095 & 0.1   & 0.1   & 0.10 \\
          & \multicolumn{1}{l}{\textbf{GL10}} & 0.08  & 0.3   & 0.5   & 0.32 \\
          & \multicolumn{1}{l}{\textbf{COS5}} & 3.12  & 1.8   & 5.0   & 90.1 \\
    \midrule
    \multirow{4}[2]{*}{\begin{sideways}\textbf{Synthetic}\end{sideways}} & \multicolumn{1}{l}{\textbf{REC}} & 1.100 & 14.7  & 4.1   & 5.86 \\
          & \multicolumn{1}{l}{\textbf{SREC}} & 1.14  & 11.2  & 5.7   & 4.96 \\
          & \multicolumn{1}{l}{\textbf{TRCE}} & 0.134 & 1.81  & 0.458 & 1.84 \\
          & \multicolumn{1}{l}{\textbf{BBL}} & 0.185 & 2.92  & 0.618 & 2.77 \\
    \midrule
    \multicolumn{6}{c}{\textbf{Geometric Mean}} \\
    \midrule
    \multirow{5}[2]{*}{} & \multicolumn{1}{l}{\textbf{Social}} & 0.058 & 0.034 & 0.059 & 8.68 \\
          & \multicolumn{1}{l}{\textbf{Web}} & 1.46  & 0.956 & -     & 112 \\
          & \multicolumn{1}{l}{\textbf{Road}} & 0.191 & 3.67  & 0.886 & 3.95 \\
          & \multicolumn{1}{l}{\boldmath{}\textbf{$k$NN}\unboldmath{}} & 0.215 & 0.287 & 0.517 & 0.603 \\
          & \multicolumn{1}{l}{\textbf{Synthetic}} & 0.420 & 5.43  & 1.60  & 3.49 \\
    \bottomrule
    \end{tabular}%
          \caption{Running time of all tested algorithms on breadth-first search (BFS). 
      The parallel implementations include PASGAL (this paper), GBBS~\cite{gbbs2021}, and GAPBS~\cite{beamer2015gap}.
      %Multistep does not support the three largest graph because the number of vertices in CW, HL14, and HL12 are larger than 32-bit integers. 
      The sequential baseline is the standard algorithm based on maintaining the visited vertices in a queue. }
  \label{tab:bfstime}%
\end{table}% 
\clearpage

\end{document}